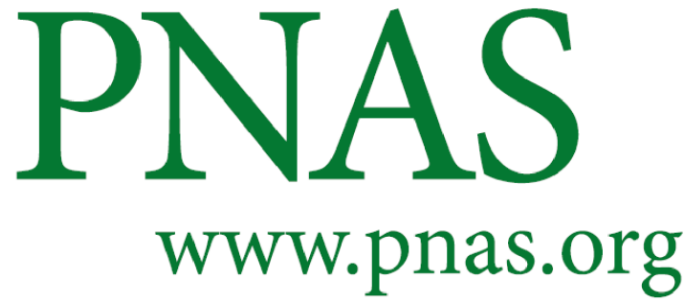

**Main Manuscript for**

**Metallic surface states in a correlated d-electron topological Kondo insulator candidate FeSb$_2$**

Ke-Jun Xu[a,b,c], Su-Di Chen[a,b,c], Yu He[a,b,c], Junfeng He[a,b], Shujie Tang[a,b,d], Chunjing Jia[a],

Eric Yue Ma[b,c], Sung-Kwan Mo[d], Dong-Hui Lu[e], Makoto Hashimoto[e], Thomas P.

Devereaux[a,b,g] and Zhi-Xun Shen[a,b,c,f,1]

[a] Stanford Institute for Materials and Energy Sciences, SLAC National Accelerator Laboratory, 2575 Sand Hill Road, Menlo Park, CA 94025

[b] Geballe Laboratory for Advanced Materials, Stanford, CA 94305

[c] Department of Applied Physics, Stanford University, Stanford, CA 94305

[d] Advanced Light Source, Lawrence Berkeley National Laboratory, Berkeley, CA 94720

[e] Stanford Synchrotron Radiation Lightsource, SLAC National Accelerator Laboratory, 2575 Sand Hill Road, Menlo Park, CA 94025

[f] Department of Physics, Stanford University, Stanford, CA 94305

[g] Department of Material Science and Engineering, Stanford University, Stanford, CA 94305

[1] Corresponding author: Zhi-Xun Shen

**Email:** zxshen@stanford.edu

**Classification**

Physical sciences: Physics

**Keywords**

Kondo insulator, strongly correlated electrons, correlated topological insulator, angle-resolved photoemission

**Author Contributions**

Author contributions: K.-J.X. and Z.-X.S. designed research; K.-J.X., S.-D.C., J.H., S.T., S.-K.M., D.-H.L., and M.H. performed ARPES measurements, K.-J.X. and E.Y.M. performed transport measurements; K.-J.X., S.-D.C., Y.H., M.H., and Z.-X.S. analyzed data; C.J. and T.P.D. provided theory support; Z.-X.S. was responsible for overall project management; and K.-J.X., S.-D.C., M.H., and Z.-X.S. co-wrote the paper with input from all authors.

**This PDF file includes:**

>Main Text
>Figures 1 to 4
>Table 1

**Abstract**

The resistance of a conventional insulator diverges as temperature approaches zero. The peculiar low temperature resistivity saturation in the 4f Kondo insulator (KI) $SmB_6$ has spurred proposals of a correlation-driven topological Kondo insulator (TKI) with exotic ground states. However, the scarcity of model TKI material families leaves difficulties in disentangling key ingredients from irrelevant details. Here we use angle-resolved photoemission spectroscopy (ARPES) to study $FeSb_2$, a correlated d-electron KI candidate that also exhibits a low temperature resistivity saturation. On the (010) surface, we find a rich assemblage of metallic states with two-dimensional dispersion. Measurements of the bulk band structure reveal band renormalization, a large temperature-dependent band shift, and flat spectral features along certain high symmetry directions, providing spectroscopic evidence for strong correlations. Our observations suggest that exotic insulating states resembling those in $SmB_6$ and $YbB_{12}$ may also exist in systems with d instead of f electrons.

**Significance Statement**

The correlation-driven topological insulator is a poorly-understood state of matter where topological protection is afforded in the absence of well-defined quasiparticles. Ongoing research is hampered by the scarcity of model material families, consisting of only the 4f rare earth boride compounds thus far. Here we establish a new class of candidate topological Kondo insulator in $FeSb_2$, based on 3d electrons instead of 4f electrons. Comparison between the different families will allow identification of key ingredients responsible for the exotic insulating state. Our band structure measurements will serve as important guides for theoretical modeling of the correlated insulating state in this system.

**Main Text**

**Introduction**

The search for correlation-driven topological physics has been focused on 4f electron Kondo insulators such as $SmB_6$ (1), where the gap opened by hybridization between localized 4f electrons and conducting 5d electrons is predicted to host topological states protected by orbital symmetry (2). The ubiquitous low temperature resistivity plateau led to proposals of surface-dominated conductivity and observations of non-local transport (3,4,5). Surface states have been found in Angle-resolved photoemission spectroscopy (ARPES) measurements (6,7,8,9,10), but the topological origin of these states is under debate (11). Furthermore, discovery of anomalous quantum oscillations (QOs) (12,13,14,15) in the insulating state suggests the existence of exotic ground states, and questions the traditional dichotomy between insulators and metals. Whether these QOs are generic properties of narrow gap insulators or only of correlated topological insulators remains a central question to this new paradigm shift. However, recent progress in studying these 4f Kondo insulators has slowed due to the low energy scale (<10 meV), difficulties in modeling the strong correlations, and the scarcity of suitable Kondo insulator material families.

We turn attention to the 3d electron narrow gap insulator $FeSb_2$, that had been previously studied mainly due to its exceptionally large thermopower at low temperatures (16,17,18,19). The physical properties of $FeSb_2$ present many striking similarities to the exotic insulating state in $SmB_6$. Firstly, electrical transport shows insulating behavior with an activation gap of ~25 meV. Below about 6 K a resistivity plateau develops, ubiquitous to samples from different synthesis method (20,21), but previously attributed to impurity states. Secondly, the magnetic susceptibility has a broad maximum at around room temperature and vanishes at low temperatures (22), suggesting the presence of spin screening and formation of Kondo singlets. Thirdly, X-ray absorption spectroscopy shows the Fe atom has a valence between $Fe^{2+}$ and $Fe^{3+}$ states (23), resembling the mixed valent $Sm^{2+/3+}$ in $SmB_6$ (24). Given these intriguing macroscopic properties, electronic band structure data are needed to understand the microscopic mechanisms. The larger gap size and energy scales here compared to the 4f compounds allows meV-resolution spectroscopic tools such as ARPES to identify the bulk and surface states that contribute to the strong correlations. We report ARPES data on the (010) cleavage plane, revealing spectroscopic evidence for metallic surface states. Further measurements of the bulk bands reveal renormalization and peculiar flat bands along certain high symmetry directions, providing corroborating spectroscopic evidence for strong correlations found in earlier transport and thermodynamic studies (22,25).

**Results**

FeSb$_2$ crystalizes in an orthorhombic marcasite structure (Fig. 1A) with lattice constants a=5.82Å, b=6.52Å, and c=3.19Å (20). The bulk Brillouin zone (BZ) and the (010) surface BZ are shown in Fig. 1B. Fig. 1C shows the resistivity and magnetic moment as a function of temperature. Below around 6 K a saturation in resistivity is seen (Fig. 1C), consistent with previous studies. The shoulder feature at around 10 K is also a reproducible feature from previous works (22). The magnetization below room temperature decreases and becomes close to zero at low temperatures (Fig. 1C), consistent with Kondo screening of the single ion moment. The slight diamagnetism is consistent with previous works (20).

Given the low temperature resistance saturation in FeSb$_2$, we looked for a Fermi surface in the insulating state at low temperatures. The Fermi energy intensity mapping at 15 K in the Y-S-R-T $k_x$-$k_z$ plane is presented in Fig. 2A. Several Fermi surfaces can be readily identified: there is an open Fermi surface along the $k_x$ direction, a closed electron pocket centered at the $\bar{\Gamma}$ point, and a closed hole pocket centered at the $\bar{Z}$ point. The 3 Fermi surfaces are labeled as α，β, and γ, as shown in Fig. 2B, where the Fermi momentum $k_F$ of each surface are identified by fitting the momentum distribution curves near the Fermi level ($E_F$). Details of the zone center pocket is shown in a high-resolution mapping in Fig. 2A enclosed by the black rectangle centered at $k_x$=$k_z$=0, revealing an apparent splitting along the $k_z$ direction. There are additional lines of intensity along the $k_x$ direction, marked by the brown arrows in Fig. 2A, that forms an open pocket (see details of this pocket in SI appendix, section S3).

We examine the 3 main Fermi surface features more closely. The β pocket spitting is shown in energy-momentum cuts in Fig. 2C and 2D. The splitting is largest along $\bar{\Gamma}$-$\bar{Z}$, reaching about 30 meV, and small or non-existent along $\bar{\Gamma}$-$\bar{X}$. The other surface bands display no splitting within our experimental resolution. The small patch of intensity at $\bar{U}$ is not from a feature that crosses $E_F$, but rather from leaked intensity of the α band just below $E_F$. This is clearly seen in the cut along the $\bar{U}$-$\bar{X}$ direction (Fig. 2E), showing that the intensity near $\overline{U}$ is connected to the α band, and is likely formed from hybridization with the bulk band. Along $\overline{Z}$-$\overline{U}$ (Fig. 2F and 2G), the γ band and α band disperses from above $E_F$ into a flat intensity at $E_B$ = 100 meV. The area enclosed by the β and γ bands is about 10% and 21% of the surface BZ, respectively (see Table 1), whereas the α band is an open Fermi surface feature.

The fact that these states cross $E_F$ in a bulk insulator suggests a surface origin. To confirm, we checked their 2D nature by varying the photon energy to probe the perpendicular $k_y$ dispersion. The mapping in the y-z plane is shown in Fig. 2H, revealing dispersionless features corresponding to the β and γ pockets. Due to the smaller velocity of the α feature along this cut, and together with the presence of the additional 1D band, the α feature $k_y$ dispersion is not clearly visible. However, we also map the Fermi surface near the Γ-Z-U-X plane (see SI appendix section S4), showing exactly the same set of Fermi surface features. Thus, we conclude that these states crossing $E_F$ are 2D in nature.

We further compare the rest of the bulk band structure to previous calculations (26,27,28) that have attempted, with different modifications to local density approximation (LDA), to reproduce the transport gap of ~25 meV. The bulk bands along the high symmetry directions are overlaid with calculated band structure from a recent work (28) in Fig. 3A to 3D. While there is some agreement when the surface states are excluded, there are also a number of discrepancies. Firstly, the calculations are missing a band with a band bottom of ~350 meV at the S point. The flat band at $E_B$ = 130 meV along S-Y roughly agrees with the calculated band, but the band width is significantly smaller than that predicted by the calculations. This can be seen clearly in the energy distribution curves (EDCs) shown in Fig. 3E and 3F, where grey region highlights the narrow band width. This band is dispersive in the $k_z$ direction, and its dispersion can be tracked in Fig. 3A and SI appendix, section S6. The anisotropic flat features here indicate localization only along the $k_x$ direction, consistent with the prediction that $FeSb_2$ has d electrons that show more itinerant behavior than the 4f KIs and should be situated between a fully hybridized Kondo Insulator and an ionic band insulator (16). While it is tempting to directly assign the flat bands as the origin of the Kondo physics, we note that there is no working theory considering anisotropically localized states as Kondo spins. However, the comparison to theoretical calculations already reveal both qualitative and quantitative differences, warranting further development of models suitable for this system. In addition, we observe a strong enhancement of spectral weight along R-T at $E_B$ = 100 meV. The EDCs in Fig. 3F shows the increased spectral weight at this binding energy, with a bandwidth of less than 20 meV. However, the spectral weight quickly disappears away from the high symmetry cut (see SI appendix, section S5). As such, the nature of this flat feature is likely different than the flat band along S-Y and seems to go beyond a simple band description.

The temperature evolution of the band structure reveals additional effects that are unaccounted for in LDA. The cut near $k_F$ of the β pocket is shown in Fig. 4A for 55K (left) and 170K (right). A clear rigid shift of the bands is seen, and the shift magnitude is quantitatively extracted from the EDCs and shown in Fig. 4B, reaching about 20 meV between 6 K and 170 K. The apparent shift in the chemical potential μ of -0.12 mV/K observed here is comparable in magnitude to that in the correlated superconductor FeSe (29). We checked for gap opening, static charging effects, and surface degradation (SI appendix, sections S7-S9), to rule out these extrinsic experimental effects. As in the case of FeSe, rigid shifts can be caused by asymmetric thermal occupation of valence and conduction bands that have different effective masses, and the shift slope is equal to the seebeck coefficient if μ is in a bulk gap (30). Current calculations places the valence band top near the R point and the conduction band bottom between Γ and Z, both with similar effective masses (26,27,28). If asymmetric thermal occupation is responsible for the band shift, then additional significant renormalizations on the near-$E_F$ bands or the presence of additional flat bands is required. However, we note that many-body effects can also induce temperature-dependent band renormalizations and shifts ( 31 ), especially considering that the shift observed here cannot account for the measured colossal seebeck coefficient of up to 45 mV/K at 10 K (17,18,19). This suggests that effects other than asymmetric thermal occupation are responsible for the large thermopower, such as the recently discussed phonon drag effect in $FeSb_2$ (19). Furthermore, the large band shift observed here renders the transport activation gap of ~25 meV inaccurate. These effects must be taken into consideration for modeling this system. We note that, while becoming broadened like the bulk bands, the β surface state is still present at 170 K, above the resistivity upturn temperature range of 100-150 K associated with the coherence crossover (see Fig. 1C), but below the single ion Kondo temperature of ~350 K (22).

**Discussion**

The observation of the 2D metallic states, together with the ubiquitous low temperature resistivity saturation, raise the immediate question of whether these states originate from topological protection. This is a difficult question that has not been fully resolved between theory and experiment even in the more established KI $SmB_6$ due to difficulties in modeling the strong correlations. In $SmB_6$, the ground state of the localized f orbitals is a well-defined quartet from the cubic crystal field (32), and the degeneracy favors the strong topological insulator case with protection arising from the orbital parities of the inverted f-d orbitals (33). In $FeSb_2$, the bulk states around $E_F$ contain all of the Fe d orbital, as well as the Sb p orbital, characters (28). The orthorhombic crystal structure would split any degenerate ground state manifolds, and the rotated Sb octahedra mix the orthogonal states. Furthermore, the more itinerant nature of the 3d electrons compared to that of the 4f compounds likely render these manifolds less well-defined as atomic orbitals. A complete topological theory for $FeSb_2$ needs to consider not only the d-p hybridizations, but also possible non-trivial d manifold topology arising from the extra mirror plane in the z direction and the resulting $Z_4$ categorization (34). We further note that in the generalized theory of cuboid KIs, both weak and strong topological KI states can still be possible in a lower symmetry KI (33). Future experiments, for example spin-resolved ARPES, non-local transport, and scanning tunneling microscopy are needed to further characterize the surface and bulk states. Combined with more refined band structure calculations, we may then be able to answer the question of whether the observed surface states are topological. We note that a recent transport work has suggested the presence of low temperature surface conduction in the related compound FeSi (35), providing a hint of evidence for non-trivial conduction in a related d-electron correlated insulator.

Another pressing question is whether QOs can also exist in $FeSb_2$. Due to the diverging interpretations of QOs in the 4f systems, a comparison to $FeSb_2$ would allow identification of common features that are key ingredients for such an exotic insulating state.

Furthermore, the measurement of QOs at high magnetic fields in $SmB_6$ and $YbB_{12}$ is complicated by a field-induced gap collapse, with the transport gap closing at 85 to 100 T (36, 37) and 50 T (38) respectively. For $FeSb_2$, the transport gap is larger than the 4f compounds, and less likely to be prone to field-induced gap-closing (see SI appendix, section S10). Thus, $FeSb_2$ would be a cleaner test case for quantum oscillations in narrow-gap correlated insulators. We also note an apparent absence of a linear component in the low temperature specific heat (39), thus observation of quantum oscillations in $FeSb_2$ would rule out theories based on charge-neutral Fermi surfaces. Further studies on other strongly-correlated Kondo insulator candidates such as FeSi, $FeGa_3$, and $Fe_2VAl$ (16) can provide additional comparison to $FeSb_2$, allowing a better understanding of this new class of d electron candidate TKIs.

**Materials and Methods**

*Crystal growth and preparation*

Single crystals of $FeSb_2$ were grown with a Sb self-flux method similar to that described in (20). The electrical transport and magnetic properties were characterized in a Quantum Design Physical Property Measurement System. To prepare the samples for ARPES measurements, the crystals were glued to a copper post using H20E silver epoxy and Torr-Seal epoxy, with a ceramic top post for *in situ* cleaving. The cleaved crystals were visually inspected after the measurement to ensure the measurement positions are free of Sb flux inclusions.

*ARPES*

The crystals were cleaved *in situ* at 15 K, exposing the a-c ($k_x$-$k_z$) plane, where the highest quality data is obtained. The Fermi surface mapping and high symmetry cuts were taken at SSRL beamline 5-2 with a Scienta D80 analyzer at energy resolutions better than 20 meV and base pressures better than $4 \times 10^{-11}$ Torr, and at ALS beamline 10.0.1 with a Scienta R4000 analyzer at energy resolutions better than 25 meV and pressures better than $7 \times 10^{-11}$ Torr. The temperature dependence measurements were performed at SSRL beamline 5-4 with a base pressure of $2 \times 10^{-11}$ Torr and energy resolution of 5 meV. A small heater around the sample is used to locally vary the temperature and keep the pressure during the temperature cycling below $2.1 \times 10^{-11}$ Torr at all times. We note that the quality of the ARPES data is highly sensitive to chamber pressure, outgassing, and the amount of surface contaminants. However, this does not preclude the existence of topological surface states, as it can be the case that the topological states move deeper into the crystal after surface degradation.

**Data Availability**

**Acknowledgments**

We thank B. Moritz, E. W. Huang, S. Sebastian, Q. Si, X. Sun, and S. Raghu for useful discussions, and Y. Jiang for assistance with magnetoresistance measurements. The work at SLAC and Stanford University is supported by the US Department of Energy (DOE), Office of Basic Energy Science, Division of Materials Science and Engineering. SSRL is operated by the Office of Basic Energy Sciences, US DOE, under Contract DE-AC02-76SF00515. A portion of this work was performed at the National High Magnetic Field Laboratory, which is supported by National Science Foundation Cooperative Agreement No. DMR-1157490 and the State of Florida. K.-J. X. and Z.-X. S. acknowledge support from the Gordon and Betty Moore Foundation.

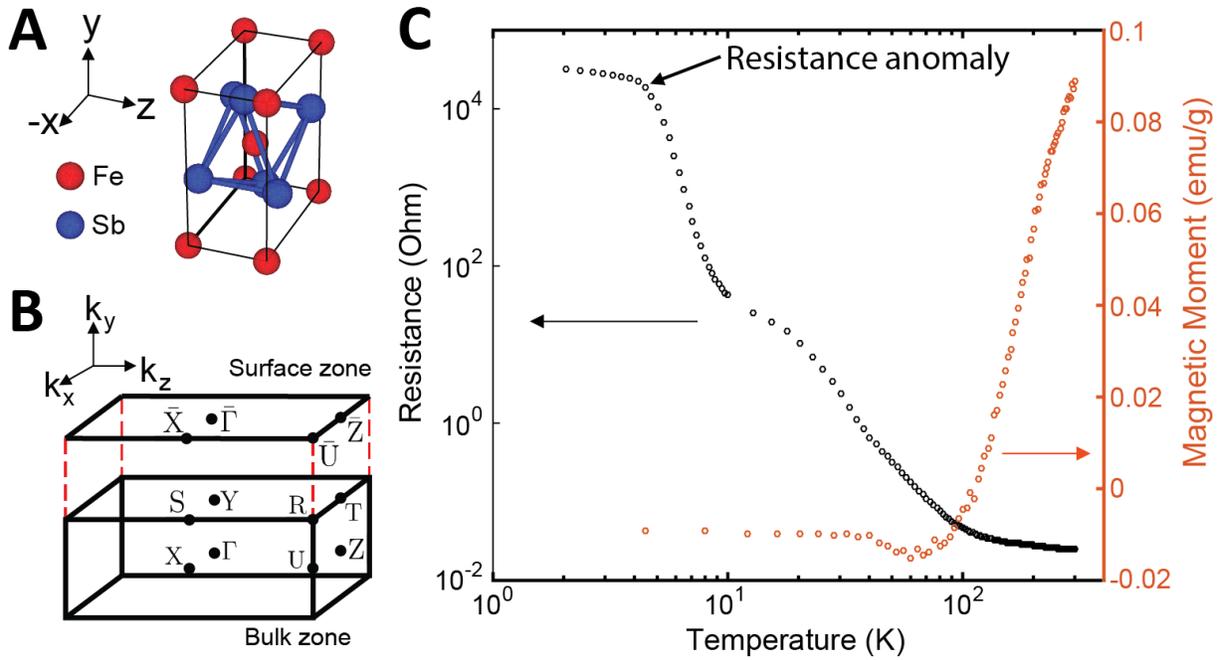

**Figure 1** Physical properties of FeSb$_2$. (A) Crystal structure of FeSb$_2$, with *Pnnm* symmetry. Fe atoms are shown in red, forming a body centered orthorhombic structure. Sb atoms, shown in blue, form a distorted octahedron around the Fe atoms. (B) Bulk Brillouin zone (BZ) and the associated surface BZ in the (010) direction. The high symmetry points in the surface BZ are labeled with a bar above the corresponding letter. (C) Resistance (left) and Magnetization (right) from 2-300 K. Below around 6 K the resistance saturates.

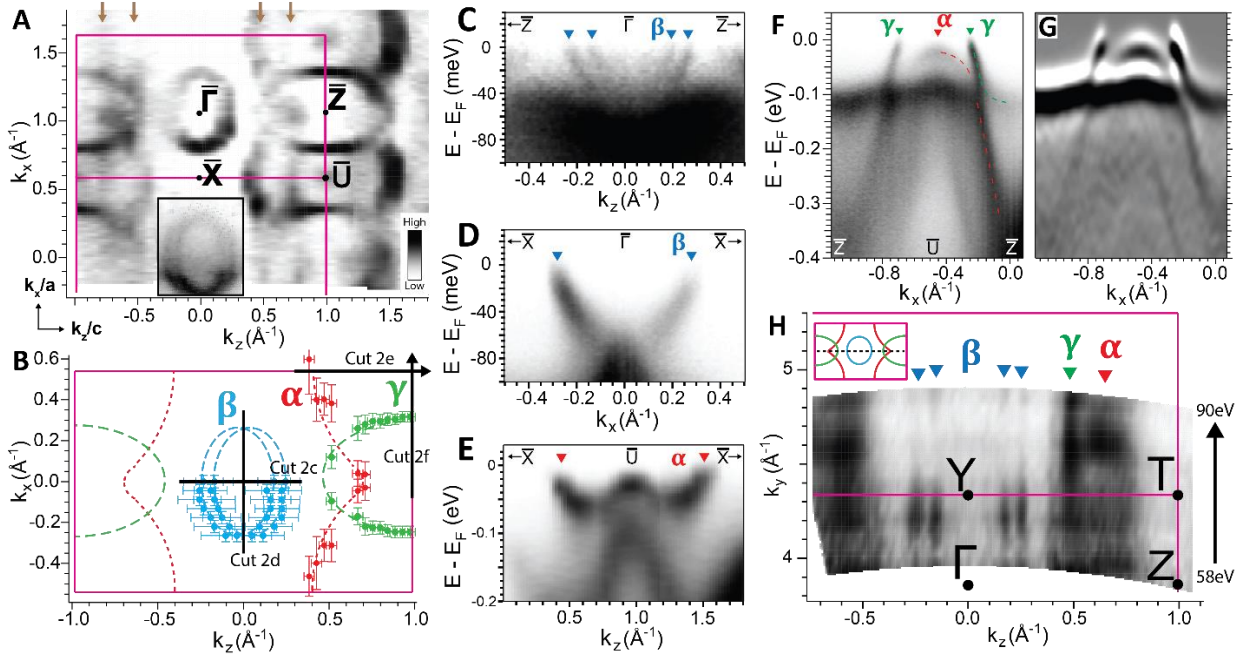

**Figure 2** Characterization of metallic surface states. (A) Fermi surface mapping in the (010) surface from spectral intensity within ±10 meV of the Fermi energy ($E_F$), taken with 70 eV Linear Horizontal (LH) photons at 15K. The surface BZ is marked with magenta rectangles and high symmetry points in black letters. Brown arrows indicate position of the additional 1D band. Region centered at (0,0) shows the zone center pocket with high momentum resolution. (B) Position of $k_F$ for the 3 pockets α, β, and γ, which are labeled in red, blue, and green respectively. $k_F$ positions are obtained from fitting the momentum distribution curves along $k_x$ for β and along $k_z$ for α and γ. The dashed lines are guides to the eye. The black solid lines indicate cut positions for the energy-momentum spectra in the rest of the figure. (C) Cut along $\bar{\Gamma}$-$\bar{Z}$, showing splitting of the β pocket. (D) Cut along $\bar{\Gamma}$-$\bar{X}$, where the β splitting is small or zero. (E) Details of the α band along $\bar{U}$-$\bar{X}$, showing that the intensity at $\bar{U}$ does not cross $E_F$. (F) Cut along $\bar{Z}$-$\bar{U}$, showing the α and γ band dispersions and the enhanced flat spectral weight at $E_B$ = 100 meV. The slight asymmetry is from the curvature of the cut. Dashed lines are guides to the eye. (G) second-derivative of the spectra in (F), showing the dispersion of the α and γ bands. (H) $k_y$ dispersion measured by varying the photon energy, using a cut along the center of the BZ as shown in the inset.

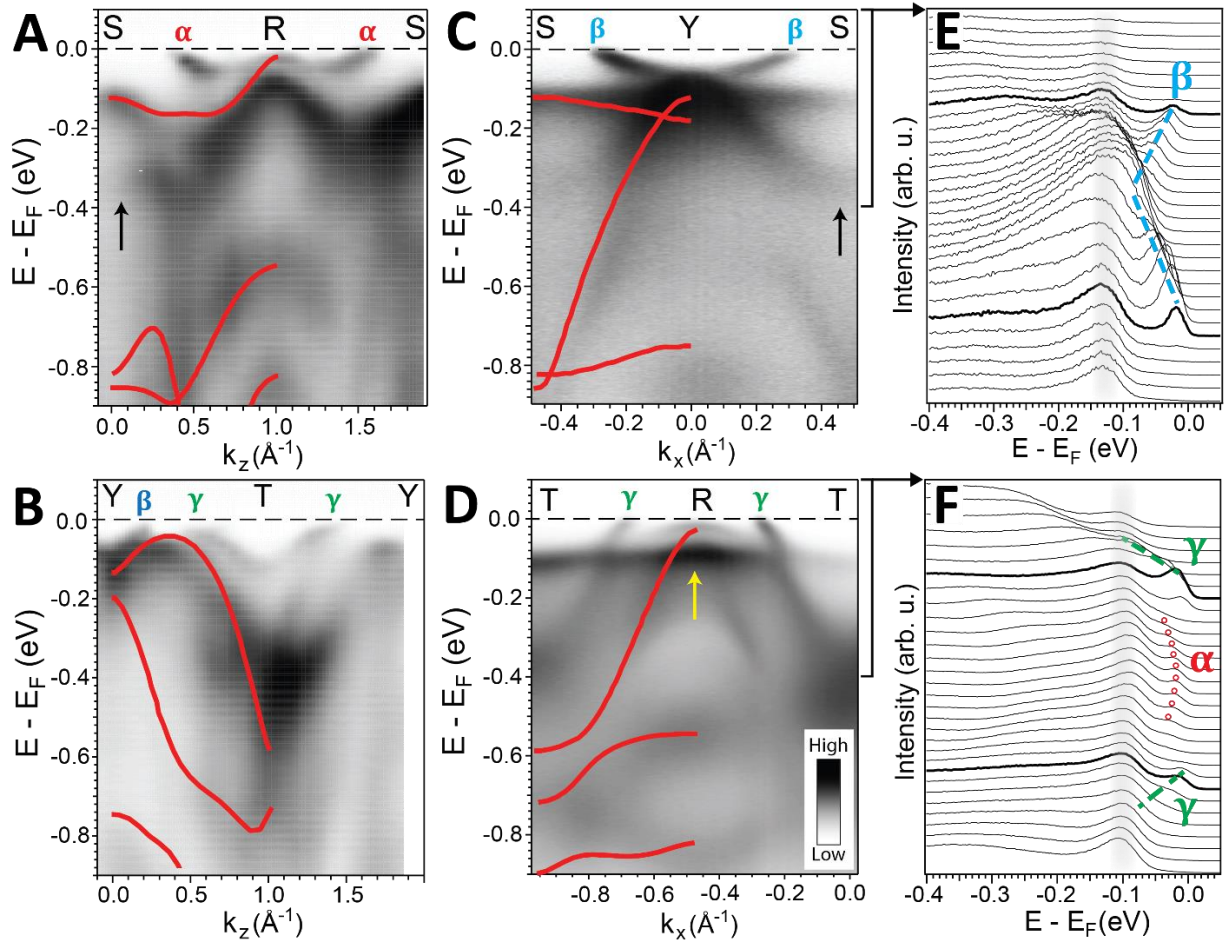

**Figure 3** Bulk bands show signatures of localization and renormalization. (A) Cut along S-R-S, (B) along Y-T-Y, (C) along S-Y-S, and (D) along T-R-T. Band structure calculations with the GW approximation from Ref. 28 are overlaid in red. The calculated curves are scaled by 0.77 to fit the experimental bulk band dispersions. A band at the S point is missing in the calculations, indicated by the black arrows in (A) and (C). The flat band along R-T is clearly visible in (D), indicated by the yellow arrow. (E) Energy distribution curves (EDCs) of the energy region enclosed by the bracket in (C), with bolded EDCs indicating $k_F$ momenta of the β pocket. Grey region highlights the flat spectral intensity at $E_B \sim 130$ meV. Dashed blue lines are guides to the eye tracing the β surface band. (F) EDCs of the bracketed region in (D). Bolded EDCs are the $k_F$ momenta of the γ pocket. Grey region highlights the spectral weight enhancement at $E_B \sim 100$ meV. Red circles and dashed green line are guides to eye tracing the α and γ surface features.

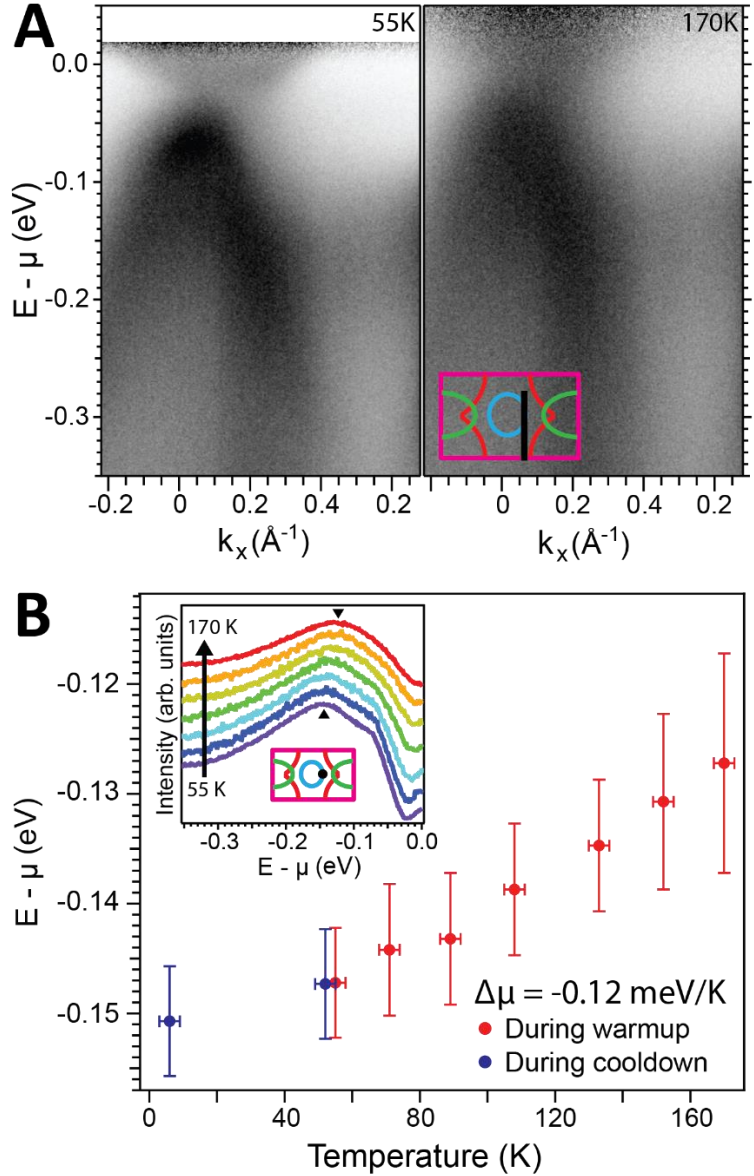

**Figure 4** Temperature dependence reveals a rigid band shift. (A) Cut near $k_F$ of the β band (see inset) taken with 21eV photons at 55 K (left) and at 170 K (right). Spectra are divided by the Fermi-Dirac distribution. (B) Position of the EDC peak (see inset) near $k_F$ of the β pocket as a function of temperature. Sample is cleaved at 55K, red data points are taken during warm up and blue points during cooldown (also see raw spectra for before and after temperature cycling in SI appendix, section I). Inset: raw EDC curves at $k_F$ of the β pocket, as shown in the BZ schematic. Black triangles indicate position of the peak corresponding to the intensity at near 150 meV $E_B$. Here μ is used to represent the general finite temperature chemical potential, whereas elsewhere $E_F$ is used for the lowest temperature spectra to indicate the zero temperature filling level.

| Band | Area enclosed by pocket (Å$^{-2}$) | % of BZ | $v_F$ (m/s) |
|---|---|---|---|
| α | N/A | N/A | 6.1×10$^3$ @ $\bar{U}$-$\bar{X}$ <br> 3.0×10$^4$ @ $\bar{\Gamma}$-$\bar{Z}$ |
| β | 0.22±0.02 | 10% | 7.5×10$^3$ @ $\bar{\Gamma}$-$\bar{X}$ |
| γ | 0.46±0.03 | 21% | 3.1×10$^4$ @ $\bar{Z}$-$\bar{U}$ |

**Table 1** Momentum space information of the surface bands. The α band is an open fermi surface and does not enclose an area. The area value for β is the average of its two component pockets.

*Supplementary Information for*

**Metallic surface states in a correlated d-electron topological Kondo insulator candidate FeSb$_2$**

Ke-Jun Xu[a,b,c], Su-Di Chen[a,b,c], Yu He[a,b,c], Junfeng He[a,b], Shujie Tang[a,b,d], Chunjing Jia[a], Eric Yue Ma[b,c], Sung-Kwan Mo[d], Dong-Hui Lu[e], Makoto Hashimoto[e], Thomas. P. Devereaux[a,b,g] and Zhi-Xun Shen[a,b,c,f,1]

[a] Stanford Institute for Materials and Energy Sciences, SLAC National Accelerator Laboratory, 2575 Sand Hill Road, Menlo Park, CA 94025

[b] Geballe Laboratory for Advanced Materials, Stanford, CA 9430

[c] Department of Applied Physics, Stanford University, Stanford, CA 94305

[d] Advanced Light Source, Lawrence Berkeley National Laboratory, Berkeley, CA 94720

[e] Stanford Synchrotron Radiation Lightsource, SLAC National Accelerator Laboratory, 2575 Sand Hill Road, Menlo Park, CA 94025

[f] Department of Physics, Stanford University, Stanford, CA 94305

[g] Department of Material Science and Engineering, Stanford University, Stanford, CA 94305

[1] Corresponding author: Zhi-Xun Shen

**Email:** zxshen@stanford.edu

## A. Cuts through the Brillouin Zone at Different Photon Energies

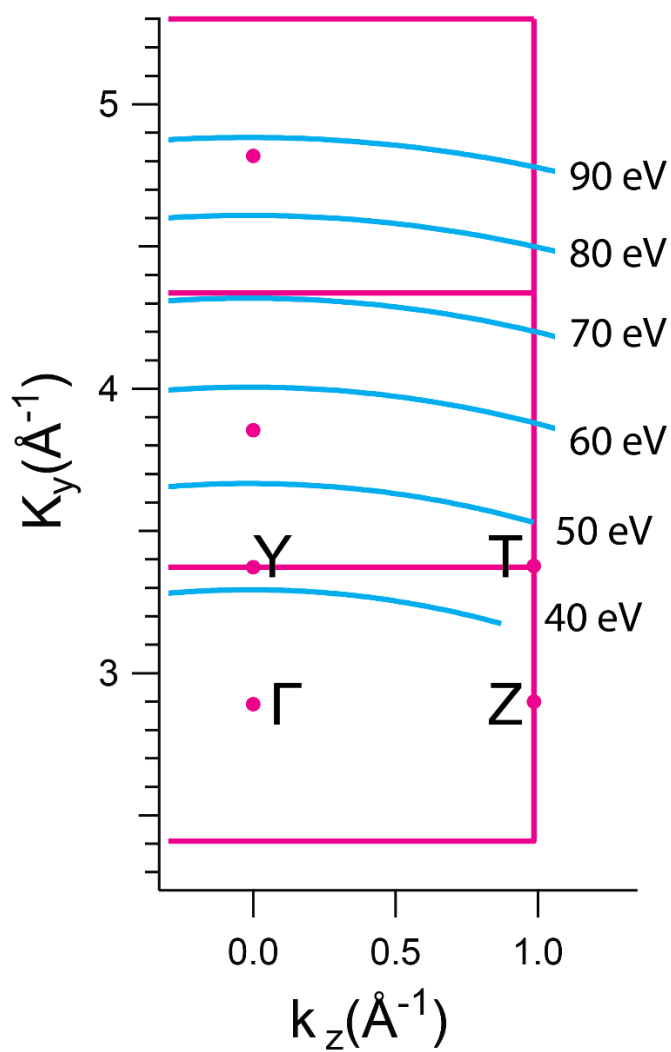

**Figure SI.A**: Calculated photon energy dependence in the perpendicular $k_y$ direction using the inner potential found in supplementary section S2, showing the locations of the cuts (blue curves) at the corresponding photon energies. The Brillouin zone in the $k_y$-$k_z$ plane is labeled with magenta rectangles, and high symmetry points are labeled with black letters. 70 eV photons cuts near the Y-T plane, while 57 eV photons cuts near the Γ-Z plane.

## B. Determination of innerpotential and the $k_y$ Brillouin zones

The out of plane momentum is determined in the following manner:

$$k_\perp = \sqrt{\frac{2m_e^*}{\hbar^2}(E_k + V_0) - k_\parallel^2} = \sqrt{\frac{2m_e^*}{\hbar^2}(E_k + V_0) - \frac{2m_e^* E_k}{\hbar^2}\sin^2\theta}$$

Where $E_k$ is the emitted electron kinetic energy, $m_e^*$ is the final state band effective electron mass, and $V_0$ is the inner potential, a parameter for the energy difference between the final state to the vacuum level. $V_0$ is usually determined in ARPES measurements by measuring the photon energy dependence of the electronic structure, then using the periodicity of the high symmetry points to extrapolate the value.

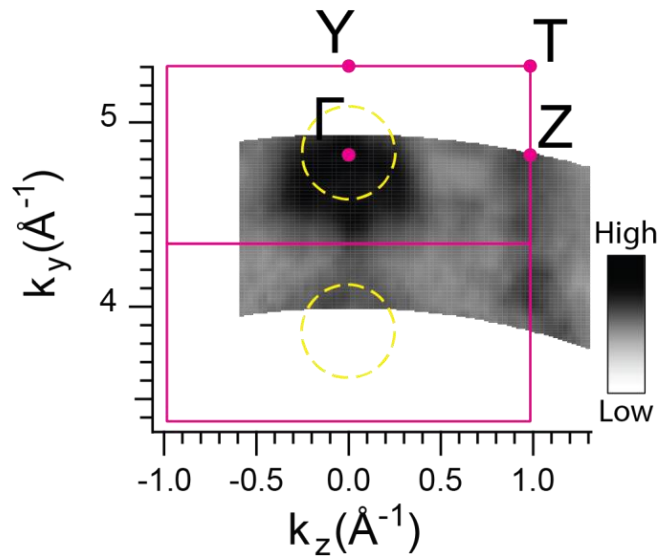

**Figure SI.B**: Determination of the inner potential. Shown here is a $k_y$-$k_z$ mapping at a binding energy of 150 meV, showing periodic features of the bulk band that allows the determination of the inner potential to be ~7 eV. Brillouin zone in the $k_y$-$k_z$ plane is labeled in magenta, and high symmetry points are labeled in black letters.

## C. Polarization Dependence and Existence of 1D pocket

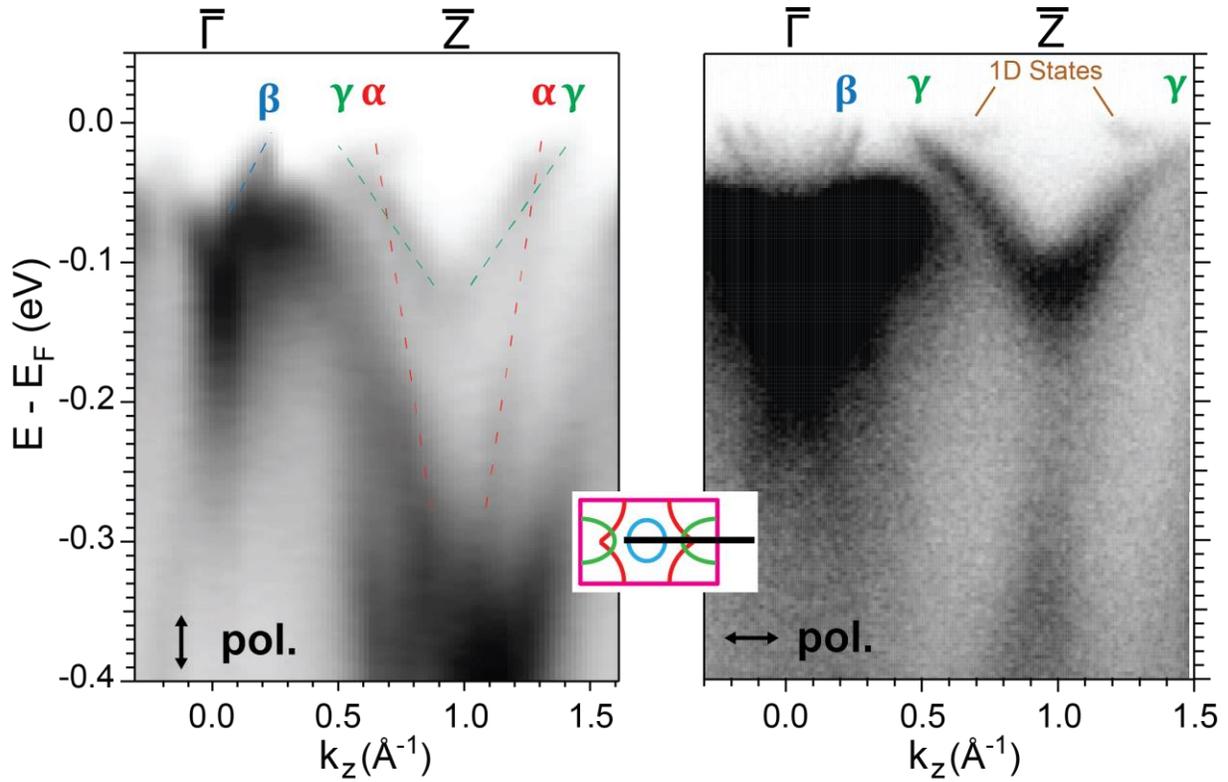

**Figure SI.C**: (left) Cut along $\bar{\Gamma}$-$\bar{Z}$ (see inset) taken with S polarization and (right) with P polarization. The arrows indicate the direction of the light for perpendicular to $k_z$ (S) and parallel to $k_z$ (P). The α band is visible with S polarization, and the band bottom at $E_B$ = 300 meV at the $\bar{Z}$ point is consistent with Fig 2f. The P polarization cut shows a shallow electron pocket that connects the 1D states (brown arrows in Fig. 2A). The S polarization spectra is a virtual cut generated from interpolating the Fermi cube from Fig. 2A.

## D. Fermi Surface Mapping at 57 eV Photon Energy

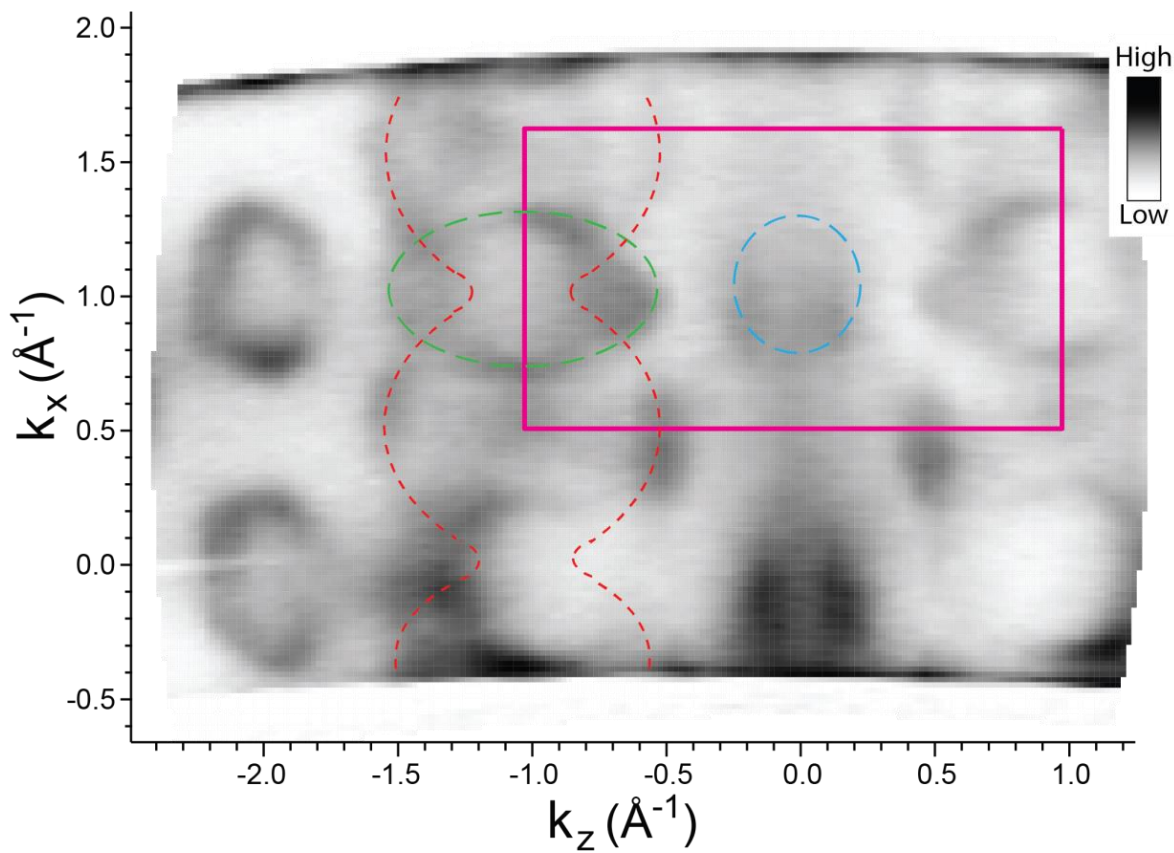

**Figure SI.D**: Fermi surface mapping taken at 57 eV photon energy, showing the same Fermi surface as that taken with 70 eV photons, even though it cuts across a different $k_\perp$ near the Γ-Z plane. BZ boundary is labeled in magenta. α band labeled in red, β band in blue, and γ band in green. Dashed lines are guides to the eye.

## E. Flat spectral weight away from high symmetry cuts

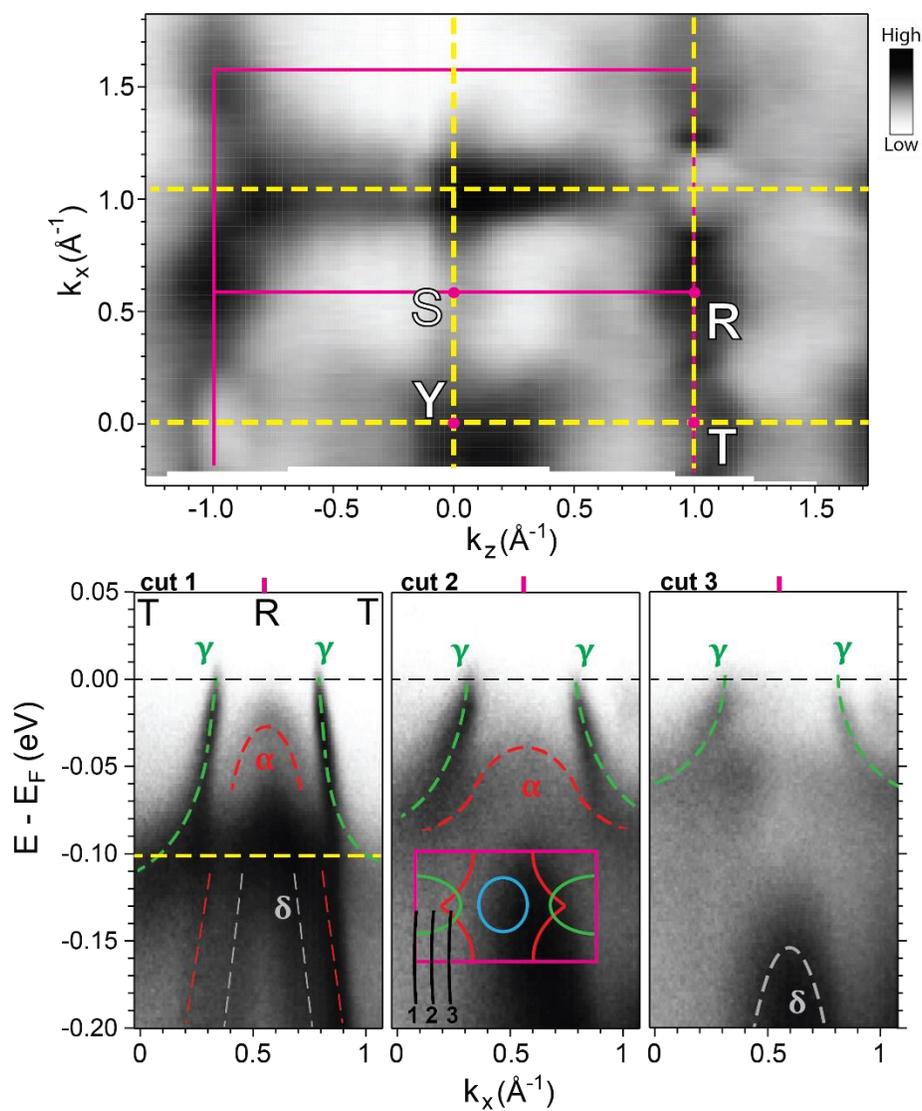

**Figure SI.E**: (top) constant energy mapping at 100 meV binding energy, showing flat intensity only along certain high symmetry directions. (bottom) cuts near the T-R-T line. The spectral weight of the flat intensity quickly disappears at the cuts away from the high symmetry line.

## F. Evolution of flat band along T-R-T

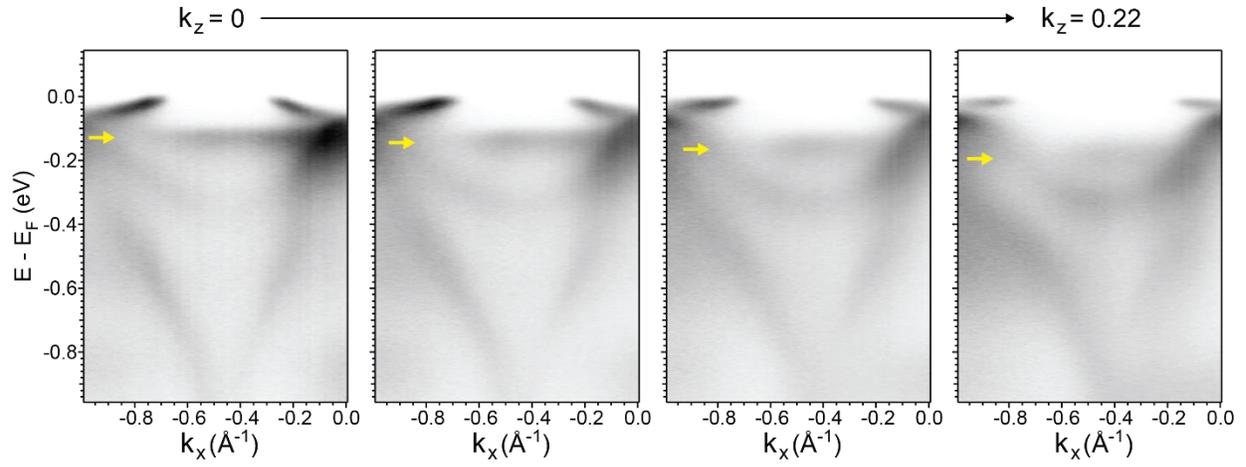

**Figure SI.F**: The flat spectral feature along T-R-T remains flat away from the high symmetry line, indicating that this is a band that disperses in the $k_z$ direction but has little to no dispersion in the $k_x$ direction.

## G. Checking for low temperature gap opening

We check for any gap that opens at $E_F$ to ensure the temperature-dependent band shift is intrinsic to the bulk states.

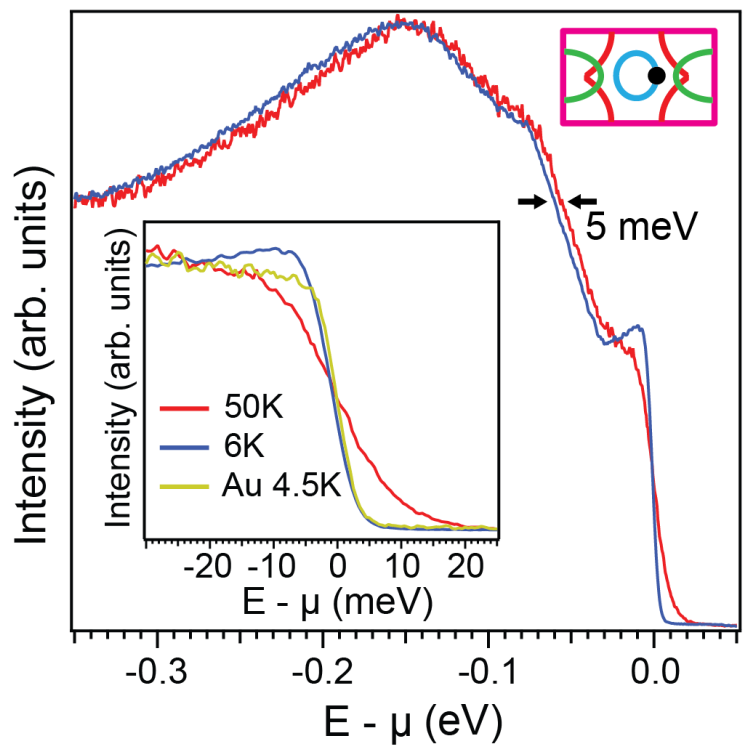

**Figure SI.G**: Comparison of the 6K and 50K EDCs near $K_F$ of the beta pocket as indicated in the BZ schematic. Inset: comparison to polycrystalline gold spectra, showing no gap opening at the lowest temperature, within the resolution of the scan of about 5 meV.

## H. Checking for charging: beam current dependence

Because FeSb$_2$ is a bulk insulator, static charging at low temperatures is a concern. Here we show, by varying the photoemission beam current, that there is minimal charging at the lowest temperatures.

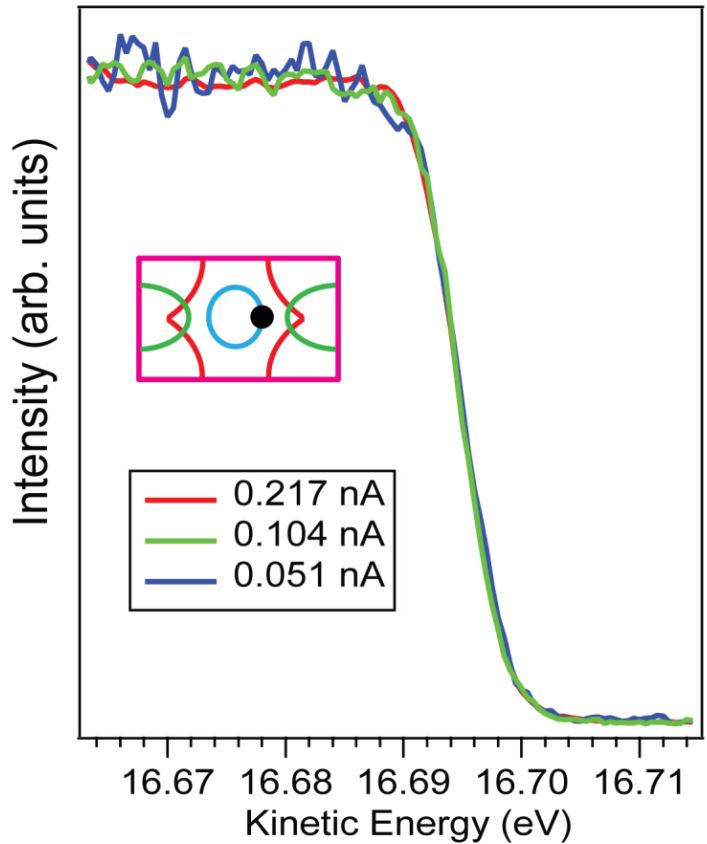

**Figure SI.H**: Beam current dependence of EDC near the β pocket k$_F$ along Y-T, the location of which is shown in the inset. The beam current was monitored at the vertical refocusing mirror, which is the last beamline optics and electrically floated. There is no leading edge shift within the resolution of the measurement. The current used here are similar or larger than that used to take the data in the main text. The lack of charging despite bulk insulation may be explained by the surface state conduction.

## I. Temperature cycling raw data

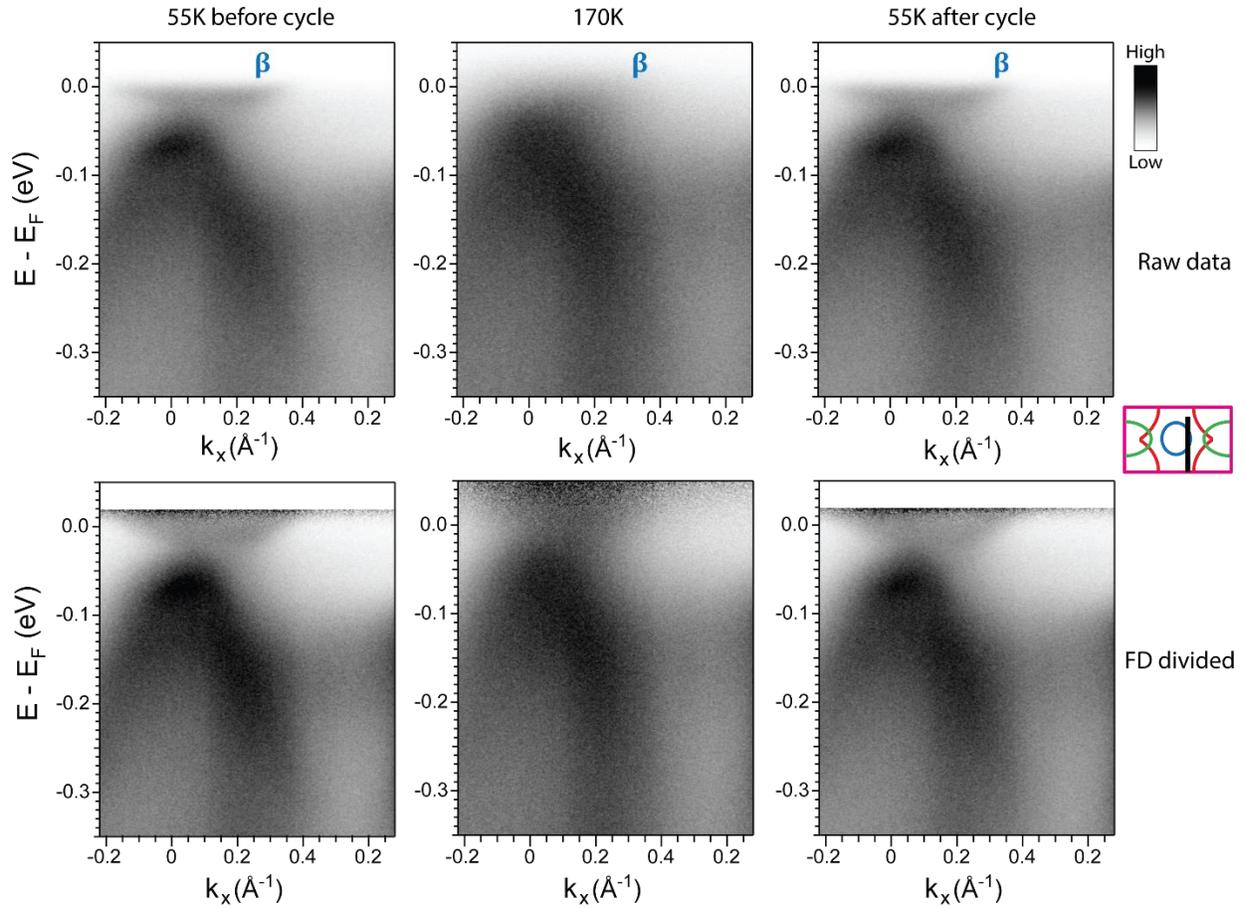

**Figure SI.I**: Raw data from the temperature cycling measurement. Top row: raw data for before and after temperature cycling. Bottom row: raw data divided by the Fermi-Dirac distribution. The cut location is shown in the inset on the side. The measurement temperatures are indicated at the top of column. The crystal is cleaved at 55K before raising temperature and cycling back to base temperature. The vacuum pressure is kept below $2.1 \times 10^{-11}$ Torr at all times during the temperature cycling. The β band, shown as the intensity crossing $E_F$, is still visible at 170K.

## J. Magnetoresistance of FeSb$_2$

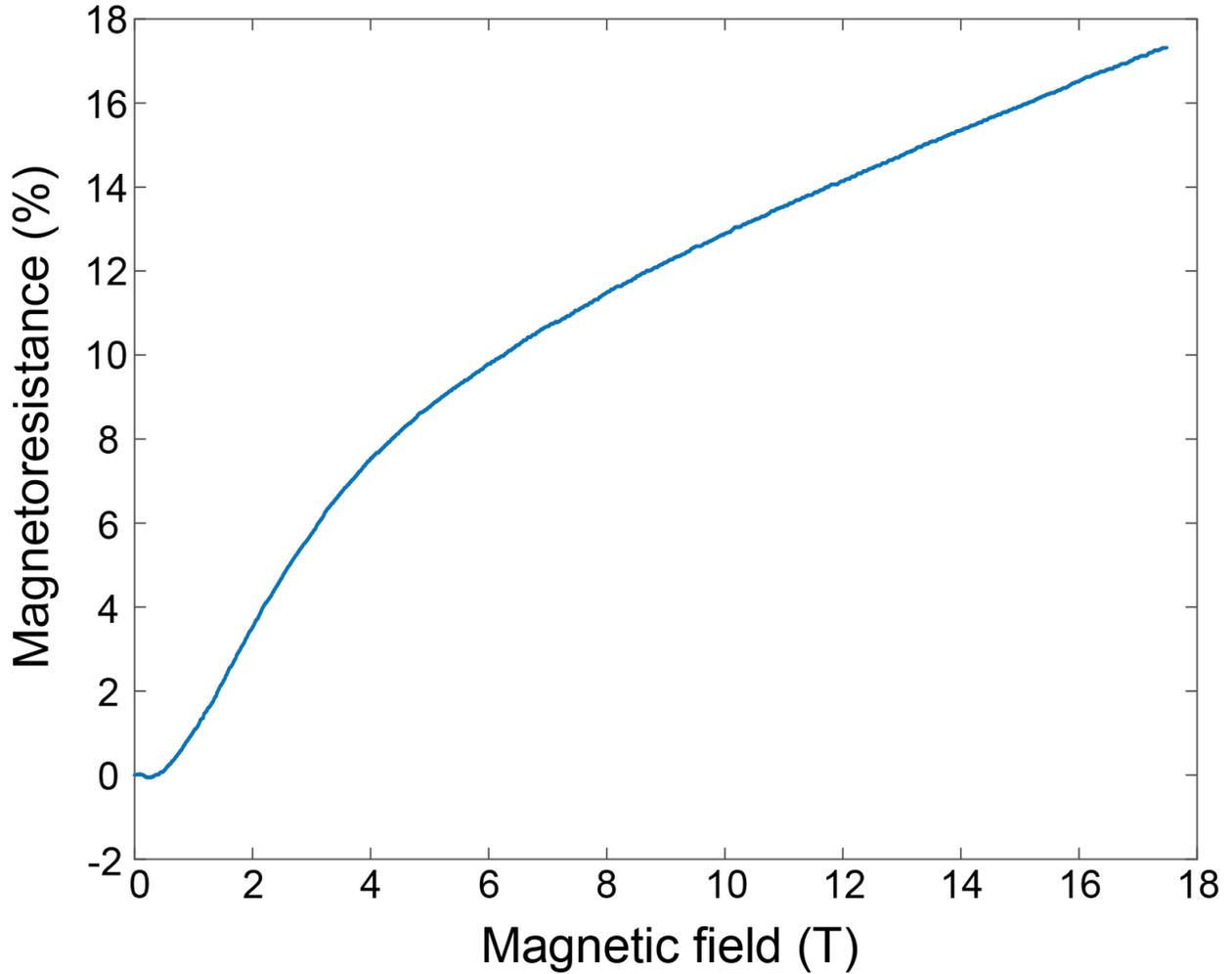

**Figure SI.J**: Magnetoresistance (MR) of FeSb$_2$ up to 17.5 T at 4.2 K, showing a positive MR for the whole field range. Data taken in the SCM3 magnet at National High Magnetic Field Laboratory using a standard 4-point geometry.